\documentclass[10pt]{article}
\newcommand{\be}{\begin{equation}}
\newcommand{\ee}{\end{equation}}
\def\bea{\begin{eqnarray}}
\def\eea{\end{eqnarray}}
\def\beas{\begin{eqnarray*}}
\def\eeas{\end{eqnarray*}}

\usepackage{latexsym}
\begin{document}

\begin{centering}

    \textbf{\Large{From the Nambu-Got\={o} to the $\sigma$-Model Action,\\
  Memoirs from Long Ago}}

  \vspace{3.5cm}

  {\large Lars Brink$^{a}$ }
  \vspace{1cm}

  \begin{minipage}{.8\textwidth}\small

\mbox{}\kern-4pt$^a$Department of Fundamental Physics, Chalmers University of Technology,
S-412 96 G\"oteborg, Sweden, lars.brink@fy.chalmers.se

\end{minipage}

\end{centering}

\vspace{4cm}

\begin{center}
  \begin{minipage}{.9\textwidth}
\textsc{Abstract}. In this article I describe my own stumblings in
the first string era. This was a time when most of the active people
were very young, not very knowledgeable and the field was completely
new. Many of us had little training for what we came to work on, and
it took quite some time to accomplish the new conceptual
discoveries.
  \end{minipage}
\end{center}

\vspace{3cm}

\centerline{\it Contribution to the volume "The Birth of String
Theory"}

\newpage

\section{Introduction}
My generation of string theorists was very fortunate. We were there
when the first ideas leading up to string theory were proposed, and
we were young and inexperienced enough not to ask too deep
questions. We could accept working in $26$ dimensions of space-time,
even when more experienced people laughed at it (and us). We were
not more clever than they were, not at all, rather we got so
attached to the ideas that we did not listen to good advice. The
average age of the active people was probably well under thirty, and
it was the last occasion where a young generation could form its
scientific future. There were a number of older heros, most notably
Yoichiro Nambu, Stanley Mandelstam, Sergio Fubini and Daniele Amati.
Also, the leading theoretical physicist of those days, Murray
Gell-Mann, was sympathetic. His words, always carefully phrased,
were listened to by all people in particle physics. This blend made
the field so exciting that once hooked it was difficult to leave the
field. After some years many had to leave in order to find
positions, but most of them had the secret wish to return to this
subject.

\section{The Formative Years}
I started as a graduate student in 1967. Sweden still had the old
system, which meant that there were no graduate schools. You had to
study on your own, and you had to work on your own. Every year the
department of theoretical physics in G\"oteborg accepted a few
graduate students, and the professors could handpick them. The first
year was spent reading books and taking oral exams once we had
finished studying a book. My advisor Jan Nilsson soon told me to
work on phenomenology, and I got in contact with an experimental
group in Stockholm, since we had no particle experimentalists in our
physics department. For some long forgotten reason, I came across
the paper by Dolen, Horn and Schmid~\cite{Dolen:1967jr} on finite-energy sum rules
during my first year, and I gave a seminar on it. I tried to follow
the subject and collected preprints but had no one to discuss it
with in our group, who mostly worked on various forms of group
theory.

I was in Stockholm in September 1968 when the professor of the
experimental group, G\"osta Ekspong, came in one day very excited
showing everybody a paper in which some Italian had found a formula
for pion-pion scattering with no free parameters other than a
coupling constant. That was the paper by Gabriele Veneziano~\cite{Veneziano:1968yb}.  Ekspong
came straight from the Vienna conference. Again I tried to follow
the subject and to study all the new concepts that appeared but back
home no one was interested. Instead I had to concentrate on my
studies of proton-proton scattering to explain the ``Deck peak,"
which was essentially the $\Delta(1236)$ resonance, and to use OPE,
which everybody knows means One-Pion-Exchange. I wrote a few papers
on it and got my licentiate degree, which is a lower doctor's
degree. After that I felt freer to study more theoretical subjects
and my advisor encouraged me to do so and, mainly together with some visitors in the department, I wrote papers on current algebra and eventually on Dual Models. These were
very simple calculations with long forgotten results, but it was a
training ground and I learned a lot. I was encouraged to apply to
CERN and, to my surprise and enormous happiness, I was accepted and
offered a fellowship from June 1, 1971, a few months earlier than
the rest of the newcomers.

\section{The CERN Years}

Life is often formed by accidental events. I came to CERN in the
beginning of the summer and met only people who were already
established in Geneva and at CERN. One month after me David Olive
came to take up a staff position leaving his job in Cambridge. We
became good friends almost immediately. We were the only two that
summer at CERN's Theory Division facing the problems that all newcomers have when they
come to CERN for a longer stay. In the beginning we were also
without our families, so we spent a lot of time together, not so
much discussing physics as discussing practical matters. David was,
of course, already famous having been one of the leaders of the
Cambridge school in the analytic S-matrix. He was one of the
old-timers (he was over 30!) who had moved into Dual Models, seeing
it as a realization of an S-matrix theory. Also during the summer I
met John Schwarz, forming a lifelong friendship; he was visiting
CERN for some weeks. I even taught him how to drive a car with a
regular gear-box. He had rented a car for a trip with his mother and
had never before driven such a car. (Many years later Chen Ning Yang
asked me if John had been my advisor and I almost said, ``no, but I
was his driving instructor.'')

When I came to CERN I was still very hesitant about what problems to
work on. I spent the summer working with some short-time visitors on
``inclusive cross sections," but I also followed all the seminars on
Dual Models. Two more lucky events happened when the new crowd
arrived at the end of the summer. One was that I got a new
officemate, Jo\"el Scherk, whom I came to share the office with for almost
two years. Jo\"el had already made a name for himself with his work
in Princeton with Andr\'e Neveu using the Jacobi imaginary
transformation to isolate the divergence in the one-loop graphs and
also with their subsequent work with John Schwarz and David Gross~\cite{Gross:1970eg}.
When he came he had just invented the ``zero-slope limit"~\cite{Scherk:1971xy}.  One of
the first days after he arrived he gave a seminar about it in the
small seminar room, and I still remember Bruno Zumino's excitement
afterwards. I overheard him say to Mary K. Gaillard that this must
have something to do with quantum field theory. (This was the
starting point for Bruno's interest in dual models and led to his
and Julius Wess's discovery of four-dimensional supersymmetry a few
years later. Bruno who had an office near ours used to come to us
and borrow all the important papers on Dual Models.) Jo\"el looked
like a genius, talked like a genius and indeed was a genius. He had
long hair and some fantastic clothes. He spoke very softly and was
always very nice to talk to. We forged a deep friendship that was
very close all the time up to his too early death in 1980 at only 33
years of age. We always had a nice discussion when he arrived in the
morning, usually about physics but often about life or Chinese
history, which he was studying on the bus to and from Geneva.

The other event that happened was that Holger Bech Nielsen
reluctantly came to CERN. He could stay for a year or more if he so
wanted. He came with his mother. When she left he stayed on in a
hotel for nine months until he went home. Holger was regarded as
{\it the} genius in the field. In his suit, which after some time
had seen better days, and his bow-tie he looked different. He could
concentrate completely on a problem; they could have dropped an
atomic bomb in the next room without disturbing him in the
slightest. He had the most remarkable ideas, which nobody else had
ever thought about. He spent all his time at CERN, eating all the
meals there and went back to town with the last bus. I am sure that
he sometimes missed it and then he walked. I had met Holger before,
and he became my entrance ticket to the Dual Model Community at
CERN. We started to work together -- mostly on his ideas. Our main
aim was to find new more realistic dual models. I did learn a lot
but our progress was not great. At some stage we used duality to get
sum rules for meson masses assuming a string with quarks at the
ends. They worked pretty well but were very sensitive to details, since we used partition functions that involved sums of
exponentials. After a year at CERN I had learned a lot but not
written any really good papers, and then Holger left.

It had, of course, been a very successful year at CERN in Dual
Models with the no-ghost theorem proved by Peter Goddard and Charles
Thorn~\cite{Goddard:1972iy}, (Charles had the third desk in our office for the year he
spent at CERN) and then their work with Claudio Rebbi and Jeffrey
Goldstone on the string~\cite{Goddard:1973qh}. There were lots of seminars and lots of
discussions. There were several collaborations going on, but by the
end of the summer David Olive and I found ourselves a bit left out.
We started to discuss and David then had the brilliant idea of
trying something really hard. (David always wants to study deep and
hard problems.) He suggested that we should try to compute the
one-loop graphs correctly. After the marvelous paper by Lovelace~\cite{Lovelace:1971fa} in
1971 on one-loop graphs, where he saw that by dividing out two
powers of the partition function and taking the space-time dimension
to be $26$, the twisted loop contains a series of poles instead of
unphysical branch-cuts, it was assumed that this should be the rule
for all one-loop graphs, but it was not proven. Nobody at that time
had a clear idea how to prove it.

This was just a year after the gauge theory revolution and my
generation and the S-matrix one, which was slightly older, knew very
little about gauge theories. We had learned QED, but our knowledge
of non-abelian gauge theories was rudimentary. The wonderful talk by
Feynman in Poland in 1963~\cite{Feynman} and the subsequent work by Faddeev and
Popov~\cite{Faddeev:1967fc} on the construction of a one-loop graphs in non-abelian gauge
theories were not known. The longer version of the Russian paper was in fact only written in
Russian. After the gauge theory revolution it was quickly translated
into English by David Gordon. David Olive and I started to study
that paper in detail as well as some marvelous lecture notes by
Abdus Salam, who as usual had immediately grasped the importance of
the subject. At the same time we studied Gerhard 't Hooft's papers.

A funny story is that the Faddeev--Popov paper discusses two
different gauge choices and concludes that they give the same
one-loop graph but not necessarily the same higher loop graphs. We
were rather mystified by that argument and asked Gerhard, who had
just arrived as a Fellow, for a discussion. Gerhard then said very
emphatically that in his paper he had shown that it worked to all
orders. No more discussion. Our problem was that we had no
Lagrangian formulation of Dual Models -- only tree diagrams. We then
devised a method in non-abelian gauge field theory of starting with
a naive one-loop graph and then deriving corrections by implementing
gauge invariance at the one-loop level. In this way we found the
ghost contributions correctly, and we then tried this method on Dual
Model loops. We worked heroically with enormous algebras, but we
could not finish it. (Many years later we realized that we had used
the wrong Virasoro generators. We had not thought of also
introducing ghosts at the two-dimensional string level.) We then
went back to square one and were told about Feynman's lecture in
Poland by Josef Honerkamp, who was a very knowledgeable field
theorist. Feynman had been interested in quantum gravity, but Murray
Gell-Mann had suggested to him  that he study non-abelian gauge
theories first as a warm-up exercise. He talked about this in the
conference. In the discussion session after the talk he was asked by
Bryce DeWitt how to compute one-loop graphs. Feynman then described
in words a method where you sew together tree-diagrams using a
projection operator onto the physical states. He said that one could
interpret the result as if two scalar ghost fields propagated
through the loop. This became the starting point for us.

Since there was no literature on this method except Feynman's words
we started by redoing it in field theory. There the projection
operator was known and easy to construct. However, for Dual Models
we had to construct such an operator. We knew that the physical
states in the critical dimension were given by the Del Giudice--Di
Vecchia--Fubini operators $ A_n^i(k)$ and their conjugates, where
$n= 1,...,\infty$ and $i= 1, ... ,d-2$. The vector $k$ is lightlike.
The projection operator could then be formally written as
\be {\cal
T}(k)= \oint \frac{dy}{2\pi i} y^{{\cal L}_0 -H- 1},
\ee
where
\be
{\cal L}_0 = \sum_{n=1}^\infty \sum_{i=1}^{d-2}
{A_n^i}^{\dagger}(k)A_n^i(k),
\ee
and
\be
 H = \sum_{n=1}^\infty \sum_{\mu=1}^{d} {\alpha_n^{\mu}}^{\dagger} {\alpha_n}_{\mu},
\ee
with $\alpha$ the ordinary harmonic oscillators of the bosonic
string, which create all the excited-string states including the
negative-norm ones.

By the use of operator product expansions and the shifting of
integration contours, we could prove the following identity for
$d=26$
\be {\cal L}_0 -H = (D_0-1)(L_0-1) + \sum_{n=1}^\infty
(D_{-n} L_n + L_{-n}D_n),
\ee
with $L_n$ the ordinary Virasoro generators and $D_n$ a new set of
operators. It is then easy to see that the projection operator is
equal to one on a physical state and zero otherwise giving us our
own proof of the no-ghost theorem~\cite{Brink:1973qm}.

With this projection operator we could set up and prove Feynman's
tree theorem in detail and then apply the same technique to the
one-loop Dual Model loops. After a lengthy calculation we could
prove that it did divide out two powers of the partition function in
the measure as Lovelace had anticipated~\cite{Brink:1973gi}.

We did not dare to send the paper to Feynman, but some months later
we got a letter from him that I still have on my office wall. He was
extremely nice to us and thanked us for writing up his theorem
``with clarity and simplicity." John Schwarz, who had moved to
Caltech at that time, had shown him the paper and he had read it
carefully.

The construction of the projection operator was like opening the
tap. We quickly redid the same calculations for the Neveu--Schwarz
model, and Corrigan and Goddard computed the projection operator for
the Ramond model. We also did the calculations for the closed-string
(or ``Pomeron'') sector at that time and proved that the
Reggeon--Pomeron vertex respected unitarity~\cite{Brink:1973nb}. We did this by
commuting the projection operator from the Reggeon (open-string)
sector through the vertex to the Pomeron (closed-string) vertex
showing that the correct projection operator appeared on that side.
This work we did with Jo\"el Scherk, who now was brought into our
collaboration. We also did the same calculation for the
fermion-emission vertex showing that indeed the Ramond and the
Neveu-Schwarz sectors were unitarily related. For this work Claudio
Rebbi also joined in~\cite{Brink:1973jd}.

All through this period I still had contact with Holger Bech. In the
spring of 1973 he came down to CERN for a week. He brought with him
a mathematical way  to compute $-1/12$ as the regularized sum of all positive integers. We thought hard how to
connect this to strings and realized quickly that by summing up all
the zero-point fluctuations of all the harmonic modes of the bosonic
string, we got just the sum of all integers. We invented a physical
way to regularize by renormalizing the velocity of light which is a
parameter of the Nambu-Got\={o} string. In this way we got an
alternative proof of $D=26$ (my third by then)~\cite{Brink:1986ja}. It was obvious to me
that the Ramond fermions must be massless since the zero-point
fluctuations canceled between the bosonic and the fermionic ones.
David, even though he is a born gentleman, persuaded me not to
publish it, since we were in the midst of our fermion calculations
and the common belief at the time was that the fermion had a
non-zero mass. I wrote it up in my thesis later that year.

At this time in the beginning of the summer of 1973 my time was up
at CERN. So I moved home to Sweden becoming very depressed. Only
when I had left did I realize what a fantastic place CERN had been
for the development of string theory. This was due to Daniele
Amati who tirelessly defended us, as we understood much later, and
who gave so much of his time and of himself to us Fellows. At home I
tried to communicate with David Olive, but this was before the
internet. Fortunately, it was also before the demise of the postal
services, so we could get letters through in less than 24 hours. The
final paper in this stage of our collaboration was the construction
of the four-fermion amplitude that David and Jo\"el constructed
during that summer~\cite{Olive:1974sv}.

Back home I also had to finish my thesis, which came to consist of
fourteen papers and an introduction of one hundred pages. It was the
longest thesis in the history of the department. After defending it
I resumed the collaboration with David Olive, and we worked hard to
understand the fermions in Dual Models. In the summer of 1974 John
Schwarz organized a workshop in Aspen and most people who had been
involved in the developments were there except for Pierre Ramond,
Charles Thorn and Holger Bech. Charles was already working on the
MIT-bag and Pierre was busy with the birth of his second daughter.
Holger's interest in string theory had started to fade, and he was
full of other interesting ideas. When I came to Aspen Bunji Sakita
told me that Professor Nambu wanted to talk to me. I was quite
excited and thought he would comment on all the work that David
Olive and I had done. No, he congratulated me instead on the paper
with Holger about zero-point fluctuations. This was very flattering,
because I consider Yoichiro Nambu to be one of the greatest
scientists of all time.

\section{Collaborations at Nordita}

I was lucky in one sense compared to my friends and collaborators.
There was no pressure on me to change to a more fashionable subject.
My situation in Sweden was stable but not very stimulating. I had
had a research position with the research council even before going
to CERN and I took it up again when I got home. It was renewed every
year; on the other hand, there were no more permanent jobs to apply
for.

When I came home after the 1974 summer at Aspen, Paolo Di Vecchia
arrived to Nordita in Copenhagen as an assistant professor. That was
to become very important for me. Nordita is a Nordic institute and
its mission is to promote theoretical physics in the Nordic
countries. I could travel to Copenhagen more or less whenever I
wanted as long as Paolo agreed, and he was also so generous that I
could stay with him when I came there. In the beginning I still
worked with Holger Bech finishing up some old ideas, but Paolo and I
discussed more and more. Paolo wanted to construct a fermionic
string by starting with $x^{\mu}$ and a space-time spinor
$\theta^\alpha$. He constructed the obvious invariant and was on his
way to construct the Superstring. I wanted to have a string action
for the Ramond--Neveu--Schwarz Model, so I was skeptical. In the ski
season of 1975 I visited CERN and had a long discussion with Bruno
Zumino. Exactly at the same time Bruno and I said that we should
have a two-dimensional spinor instead and try to have
reparametrization invariance on the world-sheet. Several people,
including Gervais and Sakita and Mandelstam, had worked earlier with
such spinors, but they had not constructed an action from which the
full constraint algebra follows. This seemed like a good problem to
work on, and I convinced Paolo to join me. We had no understanding
of Grassmann algebras and had to start from scratch. Fortunately,
there was the wonderful book by Berezin~\cite{Berezin}. We, of course, wanted to
extend the  Nambu-Got\={o} string to two-dimensional superspace but
found no way of doing it. We wanted to construct a square-root of
something but always stumbled on the strange properties of the
Grassmann variables. We also knew so little about general relativity
and the difficulty of including fermions, since neither of us had
studied any courses in General Relativity. After a while we realized
that we first ought to solve the corresponding point-particle
problem, but we ran into the same problem there.

In the summer of 1975 there was a workshop in Durham that David
Fairlie organized. That was the first time that I met Pierre Ramond.
I was already at the college where we stayed when David came in with
a new person, very French-looking. David introduced him to me and
Pierre said ``Oh, my God, Lars Brink". That was a perfect beginning
to a life-long friendship. At the meeting I realized that the
superfield formalism that we had developed was ideal for a
super-operator formalism and later that year I worked it out with my
student and formulated the superconformal formalism that later was
reinvented in the second string era~\cite{Brink:1975qk}. At the meeting there was a
crowd of Italians, all good friends of Paolo, and we started to
discuss the superfield representation we had so far for the RNS
string. We could write a free action for the superfield and then
implement the Noether current as the constraints. It worked but was
hand-waving. Soon we realized that we could extend the supersymmetry
and we all met at CERN in September to work out the $SO(2)$ case. In
this way we got an extended super-Virasoro algebra, the first one.
The key was that the Noether current also involved a Kac-Moody
$U(1)$ current. We went on and constructed an infinite sequence of
extensions~\cite{Ademollo:1975an} but only the $N=2$ and $N=4$ cases were interesting. By
using that $SO(4) = SU(2) \times SU(2)$ we could construct the
$SU(2)$ algebra. This one and the $SO(2)$ one were the only ones
with canonical operators. For the rest of that year and the
beginning of the next one we were busy formulating models for these
cases. After our first paper on the new super-Virasoro algebras I
got a letter from John Schwarz who had immediately understood our
formalism and also constructed the $SO(2)$ model. We invited him to
join us on that paper, and then the authorship consisted of eleven
Italians, one American and one Swede~\cite{Ademollo:1976pp}. Shortly after John wrote me
that Caltech had some money free for the coming academic year and
wondered if I was interested. I was already negotiating with CERN to
be a Corresponding Fellow that year, so John's offer came at a
perfect time and I accepted readily.

In the spring of 1976 we wanted to go back to the problem of finding
a reparametrization invariant action.  We slowly got to understand
the supergravity action~\cite{Freedman:1976xh} that had been constructed in the beginning
of that year. Sergio Ferrara, who had been a member of the our huge
collaboration, had moved to Paris for a year, and we should have
connected quicker to what they had done. I was very busy that spring
and summer getting another child and moving into a new house. Only
in the summer did we manage to meet and work out the particle
action. In fact, we met at CERN for a few days to finish it and were
joined by Paul Howe who was a postdoc at Nordita and by Stanley
Deser and Bruno Zumino~\cite{Brink:1976sz}. The key point, which we had missed before,
was the use of vierbeins (or, rather, as Murray Gell-Mann named
the ones in general dimensions, ``vielbeins''). Once we understood
it, it was rather straightforward to solve the particle problem and,
as a result, we got a world-line action that leads to the Dirac
equation.

We realized that we now could construct the string action, too, but
it took some time before we could meet to finalize it. One problem
for me was that I was planning the trip to Caltech and I had to make
lots of preparations for that. Finally, we met for a week and worked
like mad to construct the action and to prove all its local
symmetries and then to show that it leads to the constraints of the
RNS model~\cite{Brink:1976sc}. Again, we were inexperienced with the use of spinors and
knew nothing about Fierz rearrangements, so we had to do it the hard
way. Anyhow, I went home to get the manuscript typed by our
secretary, and then I sent it out. More or less with return mail I
received the paper by Deser and Zumino~\cite{Deser:1976rb} that contains the same action
(of course.)

The following week I left for Caltech. The first week there John and
I (very jet-lagged) constructed the $N=2$ action using the same
technique~\cite{Brink:1976vg}. It is interesting that it was this paper that Sasha
Polyakov read when he came to Caltech the next year and learned
about these actions. He had in fact been in Copenhagen for an
extended period when we constructed the action, but he was so
intensely engaged in his magnificent work on instantons and
confinement that he had missed our work. Of course, few people took
notice, since it was so far from the mainstream. We did meet each
other then briefly, though, and it was also a start of a life-long
friendship.

\section{Leaving Strings for a While}

After our first paper together at Caltech, John and I felt that we
must work on more modern stuff. We wanted to use all our insight
from string theory on supersymmetric field theories. Naturally, we
started with the ten-dimensional super Yang--Mills theory and by
compactifying it to various dimensions we found other maximally
supersymmetric gauge theories including the $N=4$ theory in $D=4$.
We got in touch with Jo\"el in Paris, who had been doing the same
thing, and we wrote it up together~\cite{Brink:1976bc}. We did feel that this was an
important model but little did we know that it should be one of the
cornerstones of modern theory. (Some five year later I returned to
it with Bengt Nilsson and Olof Lindgren when we finally found a way
to prove perturbative finiteness~\cite{Brink:1982wv}.) We wanted though to reformulate
supergravity, and we teamed up with Pierre Ramond and Murray
Gell-Mann and worked hard on supergravity in superspace. I was very
insistent on using superspace, since I had fallen in love with it in
our studies of supersymmetric strings. We worked on this for quite
some time and reconstructed the first supergravities this way~\cite{Brink:1978iv}.
Eventually in 1979 Paul Howe and I managed to construct the $N=8$
supergravity in superspace~\cite{Brink:1979nt}. When he and Ulf Lindstr\"om~\cite{Howe:1980th} using our formalism the year
after showed that there were possible counterterms in that theory I
went back to string theory and joined up with John Schwarz and
Michael Green. My attachment to strings was a love for life.

\section{Acknowledgments}

I am grateful to the organizers of "The Birth of String Theory", Andrea Cappelli, Elena Castellani, Filippo Colomo and Paolo Di Vecchia, for giving me the opportunity to present my recollections. I also wish to thank Pierre Ramond, John Schwarz and Paolo Di Vecchia for reading the manuscript and for their helpful suggestions.

\end{document}